\documentclass[aps,twocolumn,prl]{revtex4}
\usepackage{graphicx}

\begin{document}
\title{A description of the yrast states in $^{24}$Mg 
by the self-consistent 3D-cranking model}
\author{Makito Oi}
\affiliation{Department of Physics, University of Surrey, Guildford, GU2 7X, 
  Surrey, United Kingdom}

\begin{abstract}
  With the self-consistent 3D-Cranking model,
  the ground-state rotational band in $^{24}$Mg is analysed.
  A role of triaxial deformation is discussed, in particular, in
  a description of the observed two $I^{\pi}=8^+$ states.
\end{abstract}

\maketitle

\section{Introduction}
$^{24}_{12}$Mg has been studied very well
as a typical case of well-deformed light-mass nuclear systems.
After self-consistent microscopic calculations 
of the non-relativistic \cite{BFH87}
and relativistic  methods \cite{KR88} in the late 1980s,
the nucleus is believed to have an axially-symmetric 
and prolate shape in the ground state.
Assuming the core of $^{16}$O, 
the ground state configuration is supposed to have 
eight valence particles occupying the d$_{5/2}$ orbits 
(four neutrons and four protons).
Since the Fermi level of this nucleus is in the beginning of the
sd-shell (corresponding to the $N=2$ harmonic oscillator shell),
there are many open valence orbitals above the Fermi level, which
may induce deformation, and consequently a collective rotation.
Such a rotational band has been already identified in experiments
(for instance, see Ref. \cite{RB68,WWJ01}). 
A particular interest is an existence of the 
two $I^{\pi}=8^+$ states observed in experiment.
These two states are energetically close to each other 
(the difference is about 2 MeV).
After the study of Sheline et al. \cite{SRA88}, 
it is believed that the second $8^+$ belongs to the ground-state
rotational band (g-band). 
Valor, et al. analysed the g-band with the cranked Skyrme HF + BCS approach
as well as the configuration mixing approach based on the generator coordinate
method (GCM) \cite{VHB00}. 
However, their calculations assume axial symmetry for descriptions of
intrinsic states.
Up to $I^{\pi}=4^{+}$, they were 
able to reproduce the experimental data very well.
The cranked mean-field calculation gave a fairly good agreement for
$I^{\pi}=6^+$, while the configuration mixing approaches returned
larger energies for the state. This is simply because
these states at high spin 
were projected out from the non-cranked mean-field state.
Interestingly, the cranked mean-field calculation for $I^{\pi}=8^+$
matches the observed energy of the first $8^+$ state, but the authors
dismissed the agreement on the basis of the Sheline's analysis \cite{SRA88},
and they speculated that the disagreement might mainly come from 
the triaxial effects at high spin induced by the disappearance 
of the pairing correlation in their calculation.

\section{3D-cranked HFB method}
Inspired by the study by Valor et al. \cite{VHB00},
we performed the self-consistent cranking calculation 
allowing triaxial deformation in a self-consistent manner. 
As a new aspect in our study, not only 1D-cranking but also 
3D-cranking calculations were carried out.
An advantage of the 3D-cranking model 
is that low- and high-K intrinsic structures can be systematically 
studied \cite{OWA02}.

The Hamiltonian used in our study reads
\begin{equation}
  \hat{H}=\hat{H}_0 + \hat{V},
\end{equation}
where the first term describes the one-body part, which is
the spherical Nilsson Hamiltonian in this study, 
while the second part is the two-body interactions, which
is the pairing-plus-quadrupole force (so-called P$+$QQ force).
The model space (valence space) to diagonalise the two-body part
contains two major shells ($N=2,3$) in the spherical Nilsson model,
in accordance with the Kummar-Baranger criteria for the P$+$QQ force 
\cite{KB65}.
The variational state is the Hartree-Fock-Bogoliubov ansatz,
which is a generalised product state. With quasi-particle
annihilation operators $\beta_q$, the ansatz is expressed as
\begin{equation}
|\text{HFB}\rangle=\prod_q\beta_q|0\rangle,
\end{equation} 
where $|0\rangle$
is the vacuum for the canonical basis $a_m$ and $a^{\dag}_m$.
(In our case, the canonical basis correspond to the spherical Nilsson
basis.) The canonical basis and the quasi-particle basis are connected by
a unitary transformation called the general Bogoliubov transformation
\cite{On86}.
The variational equation is derived for
\begin{equation}
  \delta\langle\text{HFB}|\hat{H}-\sum_{i=1}^{3}\left(\omega_i\hat{J}_i
    +\mu_i\hat{B}_i\right)-\sum_{\tau=\text{p,n}}\lambda_{\tau}\hat{N}_{\tau}|
  \text{HFB}\rangle = 0.
\end{equation}
In the above equation,
$\hat{J}_i$ is the $i$-th component of the angular momentum operator
(the index $i$ takes $i=1,2,3$),
and $\hat{N}_{\tau}$ describes the number operator for proton ($\tau=\text{p}$)
and neutron ($\tau=\text{n}$).
$\hat{B}_i$ is an off-diagonal component of the quadrupole operator,
defined as
\begin{equation}
  \hat{B}_i = \frac{1}{2}\left(\hat{Q}_{jk}+\hat{Q}_{kj}\right),
\end{equation}
where the indices $(i,j,k)$ should be placed in a cyclic manner.
Each term with the Lagrange multipliers ($\omega_i,\mu_i$ and $\lambda_{\tau}$)
is necessary to put constraints in intrinsic states:
$(\langle\hat{J}_1\rangle,\langle\hat{J}_2\rangle,\langle\hat{J}_3\rangle)
= (J\cos\theta, J\sin\theta\sin\phi,J\sin\theta\cos\phi);
\langle\hat{N}_{\tau}\rangle =  N_{\tau}; 
\langle\hat{B}_{i}\rangle = 0.$
The last constraint is necessary so as to keep the orientation of
the angular momentum vector against the intrinsic coordinate axes \cite{On86}.
The variational equation is solved by means of the method of steepest descent.
Details of the method are presented in Ref.\cite{On86}.
A deformed Nilsson + BCS state is used for an initial trial state at $J=0$.
Deformation parameters and gap energies for the trial state are
determined by referring to the calculations of the liquid drop model 
by the M\"oller and Nix \cite{MN95}. 
In the present study, the deformation parameters for the ground state
are chosen to be $\left(\beta,\gamma)=(0.347,0.0^{\circ}\right)$ and
the pairing gap energies are $\left(\Delta_{\text{p}},\Delta_{\text{n}}\right)
=(1.840 \text{ MeV},1.962 \text{ MeV})$.
All the physical quantities, such as energy, quadrupole moments (deformation),
single-particle spin components, and gap energies, 
are self-consistently calculated
under the above constraints in this framework.

\section{Results and Discussions}

\begin{figure}
  \includegraphics[width=0.45\textwidth]{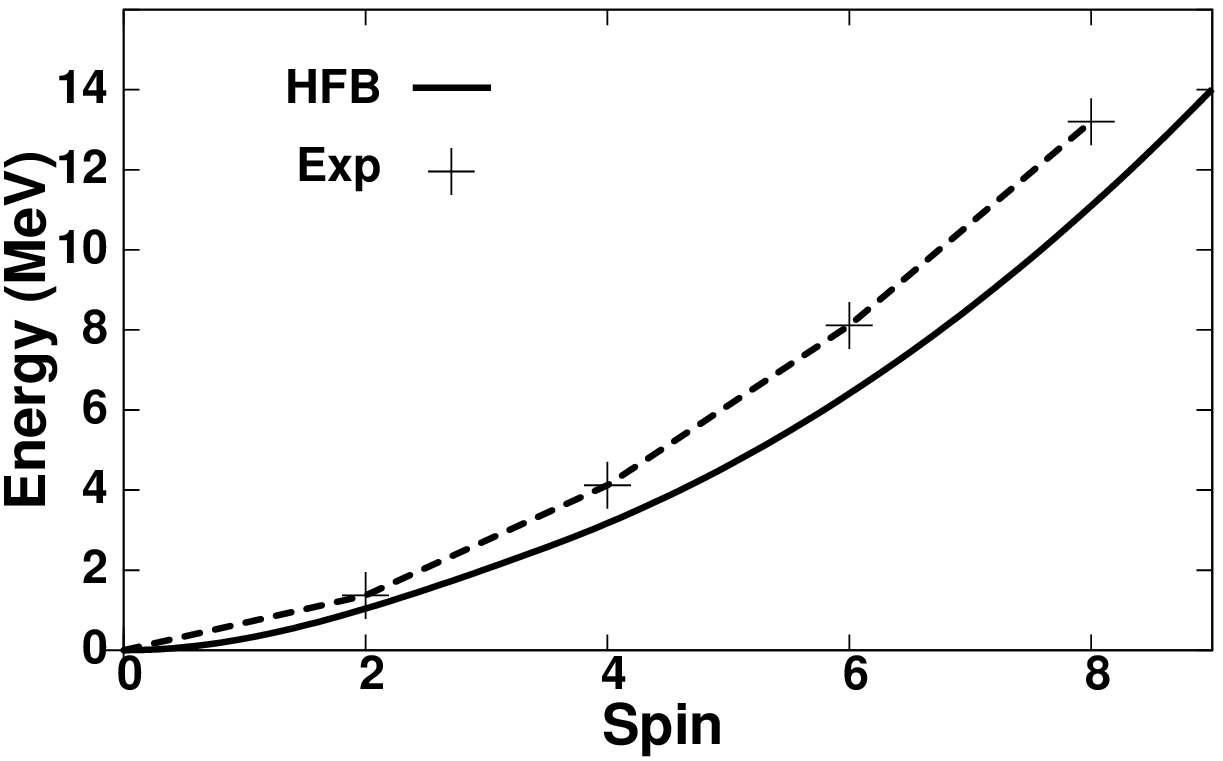}
  \includegraphics[width=0.45\textwidth]{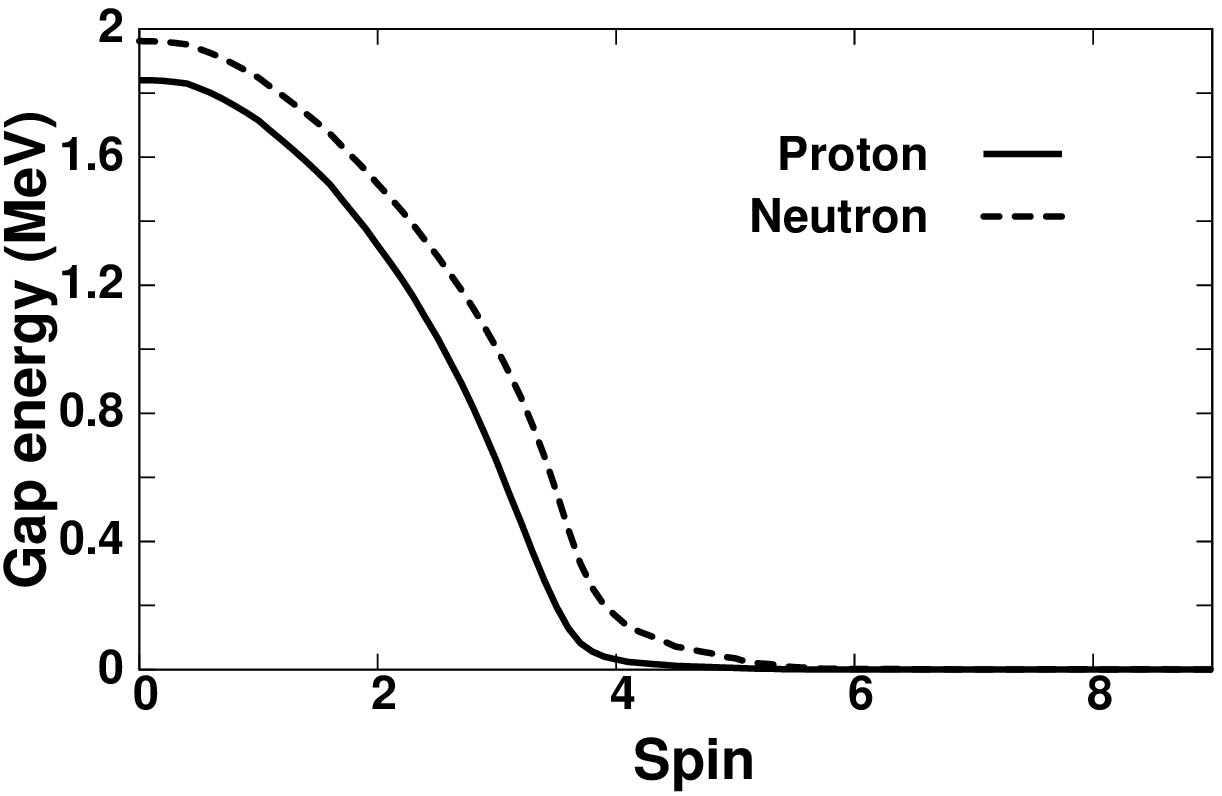}
  \includegraphics[width=0.45\textwidth]{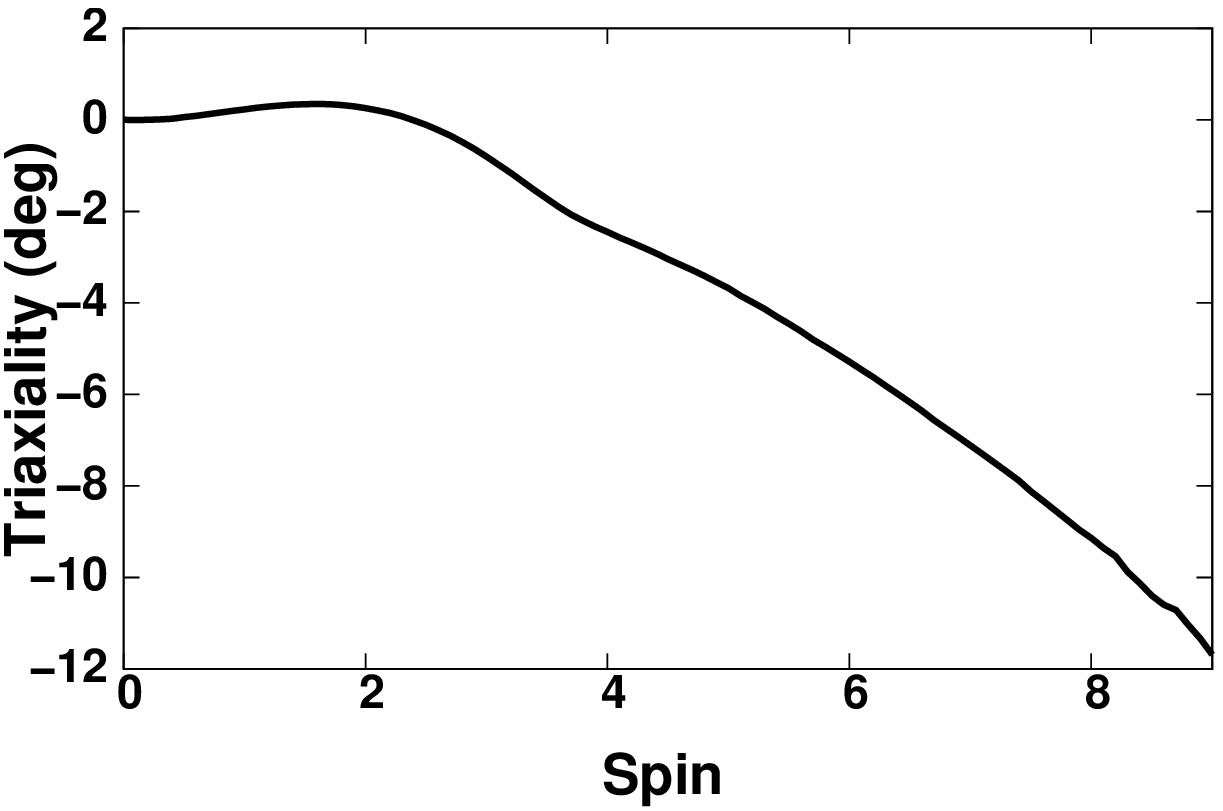}
  \caption{(top) Rotational energy obtained in the 1D cranking calculation 
    (solid line), and the experimental data (cross).
    (middle) Gap energies for protons and neutrons, obtained in the 1D cranking 
    calculation.  (bottom) Gamma deformation obtained in the 1D cranking 
    calculation.}
    \label{fig-ene}
\end{figure}

First of all, let us report the results from the 1D-cranking calculation.
Despite a use of a simple separable interaction,
the ground-state rotational spectra is reproduced reasonably well 
(Fig.\ref{fig-ene}).
As Valor, et al. commented in Ref.\cite{VHB00},
the gap energies disappear both for protons and neutrons
before $J=6\hbar$ (Fig.\ref{fig-ene}).
Triaxial deformation gradually decreases from $\gamma=0^{\circ}$.
However, in $J\le 4\hbar$, triaxial deformation can be still regarded 
negligible. In other words, axial symmetry is kept fairly well. 
On the other hand, at high spin ($J=8, 10\hbar$), 
substantial triaxial deformation is formed ($\gamma \agt -10^{\circ}$),
and axial symmetry is clearly broken. 
It should be noted here that
the convention for the quadrupole deformation parameters ($\beta,\gamma$)
in this study follows the Hill-Wheeler coordinates, which gives
the opposite sign in $\gamma$ to the so-called Lund convention.
A fact that the gamma deformation becomes negatively larger implies
that the nucleus is reaching the non-collective rotational state 
($\gamma=-60^{\circ})$, where the rotational axis corresponds 
to the shortest principal axis of the deformation.

\begin{table}[tb]
  \begin{tabular}{ccccccc}
    \hline
$J$  \quad\quad &  $0$& $2$& $4$& $6$& $8$& $9$\\
  \hline
$\beta$ \quad\quad & 0.37 & 0.38 & 0.37 & 0.35 & 0.31 & 0.29\\
\hline
  \end{tabular}
  \caption{Evolution of the deformation in $\beta$ as a function of the total spin $J$, which are obtained from the 1D-cranking calculation.}
\end{table}

From this result,
it is understandable why the cranking calculation by Valor, et al. \cite{VHB00}
was successful up to $J=6\hbar$  and why the deviations from the 
experimental values become larger at higher spins. 
As mentioned earlier, 
they solved the HFB equation within the axial symmetry constraint.

\begin{table}[htb]
\begin{tabular}{cccccc}
  \hline
  &  $J=2$& $J=4$& $J=6$& $J=8$& $J=9$\\
  \hline
  $\pi$d$_{5/2}$ &
  1.0 ($50\%$)  & 2.0 ($50\%$) &
  2.9 ($48\%$)  & 3.6 ($45\%$) &
  3.9 ($43\%$)\\
  $\pi$d$_{3/2}$ &
  0.0 ($0\%$)  & 0.0 ($0\%$) &
  0.1 ($2\%$)  & 0.4 ($5\%$) &
  0.6 ($7\%$)\\
  \hline
  $\nu$d$_{5/2}$ &
  1.0 ($50\%$)  & 2.0 ($50\%$) &
  2.9 ($48\%$)  & 3.6 ($45\%$) &
  3.9 ($43\%$)\\
  $\nu$d$_{3/2}$ &
  0.0 ($0\%$) & 0.0 ($0\%$) &
  0.1 ($2\%$) & 0.4 ($5\%$) &
  0.6 ($7\%$)\\

  \hline
\end{tabular}
\caption{Single-particle components of the total spin in the 1D cranking
calculation.The first two rows correspond to proton orbitals, while
the last two to neutron orbitals. The numbers in brackets are ratios of
single-particle spins against the total spin. The unit of spin is
$\hbar$.}
\label{tb2}
\end{table}

In the HFB theory, the total spin is expressed as the sum
of single-particle spins, that is,
\begin{equation}
  \langle\hat{J}_i\rangle
  = \sum_{m}\langle j_i^{(m)}\rangle = \sum_{mn}(j_i)_{mn}\rho_{nm},
\end{equation}
where $\rho$ is the density matrix and $j_i$ is the single-particle
angular momentum operator. The indices $m$ and $n$
are for the canonical basis, while the index $i$ for the coordinate axes,
 that is,  $i=1,2,3$. Using this information of angular momentum, 
we can discuss nuclear structure with single-particle spins.
In the Table \ref{tb2} are the calculated main components of single-particle spins
for the different total spin $J$. The result reflects a fact that 
$^{24}$Mg is a $N=Z$ nucleus, that is, the way of single-particle excitations
is the same both for protons and neutrons.
For low-spin members in the rotational band, the total spin
consists mainly of the d$_{5/2}$. The higher the total spin,
the more the d$_{3/2}$ orbit is occupied.
Therefore, the ``collectivity'' in this nucleus is attributed
to gradual excitations into the d$_{3/2}$ orbit.

Next, let us present the results of the 3D-cranking calculations.
In this paper, we focus on the analysis of the $J^{\pi}=8^{+}$ state,
where its triaxial deformation becomes substantial ($\gamma\simeq -10^{\circ}$)
in the 1D-cranking calculation.
\begin{figure}[t]
  \includegraphics[width=0.5\textwidth,angle=0]{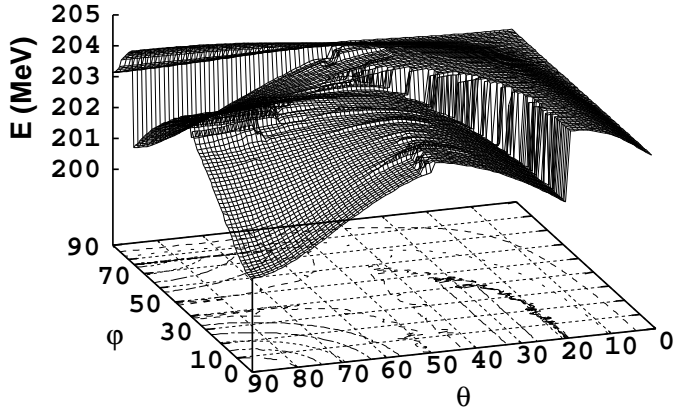}
  \caption{Energy surface of the $J=8\hbar$ state, calculated by means
  of the 3D-cranked HFB method. There are two minima as intrinsic states,
which are seen at $(\theta,\phi)=(0^{\circ},0^{\circ})$ and 
$=(90^{\circ},0^{\circ})$.}
\label{3d-ene}
\end{figure}
Obviously from the Fig.\ref{3d-ene},
two configurations compete with each other.
This competition can be considered as a kind of ``level crossing'' 
between two different states (or configurations), but
in the 3D-cranking calculation each ``level'' is represented by a curved
surface. 
There are mainly two minima in the energy surface for the $J=8\hbar$ state
(Fig.\ref{3d-ene}): $(\theta,\phi)=(0^{\circ},0^{\circ})$ and 
$=(90^{\circ},0^{\circ})$, and they characterize the two configurations.
 The former minimum corresponds to the 1D-cranking
solution where triaxiality is calculated to be $\gamma\simeq -10^{\circ}$.
The rotation axis is found to be along the shortest axis,
and the corresponding state is expected be of low-$K$ character.
The latter minimum is the energetically lowest state (yrast state) 
at this spin, which is about 2.5 MeV lower than the 1D-cranking solution.
This yrast state at $J^{\pi}=8^+$ is found to be axially symmetric because
the triaxiality is calculated to be $\gamma\simeq 0^{\circ}$
($\beta\simeq 0.24$).
In this case, the rotational axis 
points along the longest axis of the axially symmetric shape, 
so that the rotation is of single-particle character.
As a result, the major components of the state should be of high-$K$
characters.
Studying in detail the microscopic structure,
the total spin is found to be constructed almost purely by
the d$_{5/2}$ orbits (in both protons and neutrons):
$3.97\hbar$ each by the d$_{5/2}$ orbits of protons and neutrons.
In addition to the difference in the deformation,
a lack of the d$_{3/2}$ component in the first $8^+$ state
implies the yrast state is surely different from rotational members 
of the g-band, from a microscopic point of view.
The shell model calculation by Wiedenh ver, et al. \cite{WWJ01}
says that such a configuration, that is, (d$_{5/2}$)$^8$,
corresponds to the first $8^+$ state (which does not belong to
the g-band) observed in experiment. Therefore, our result is
consistent with the shell model calculation, too.

From this result, we can conclude that the yrast state found in our
calculation at $J=8\hbar$ is a high-$K$ state with $K^{\pi}=8^{+}$. 
It was experimentally observed to be 
energetically lower than the second $8^+$ state by about 2 MeV.

\begin{figure}[th]
  \includegraphics[width=0.45\textwidth]{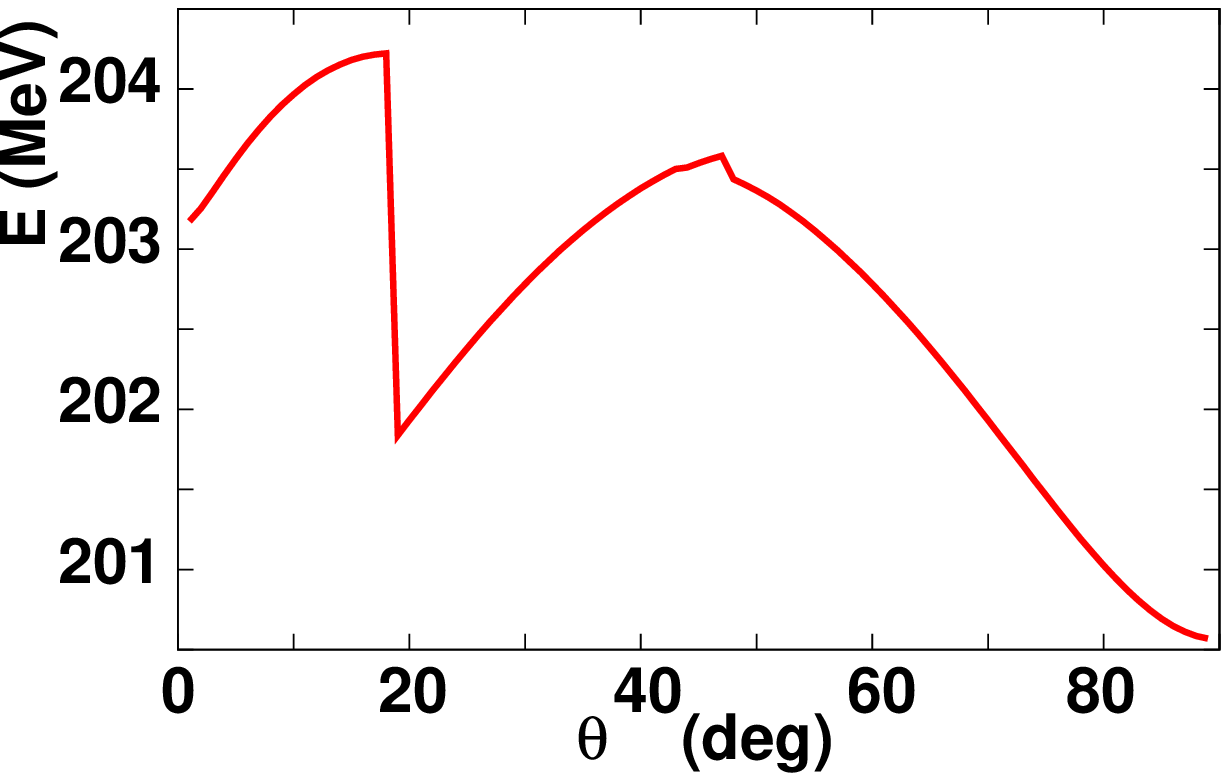}
  \includegraphics[width=0.45\textwidth]{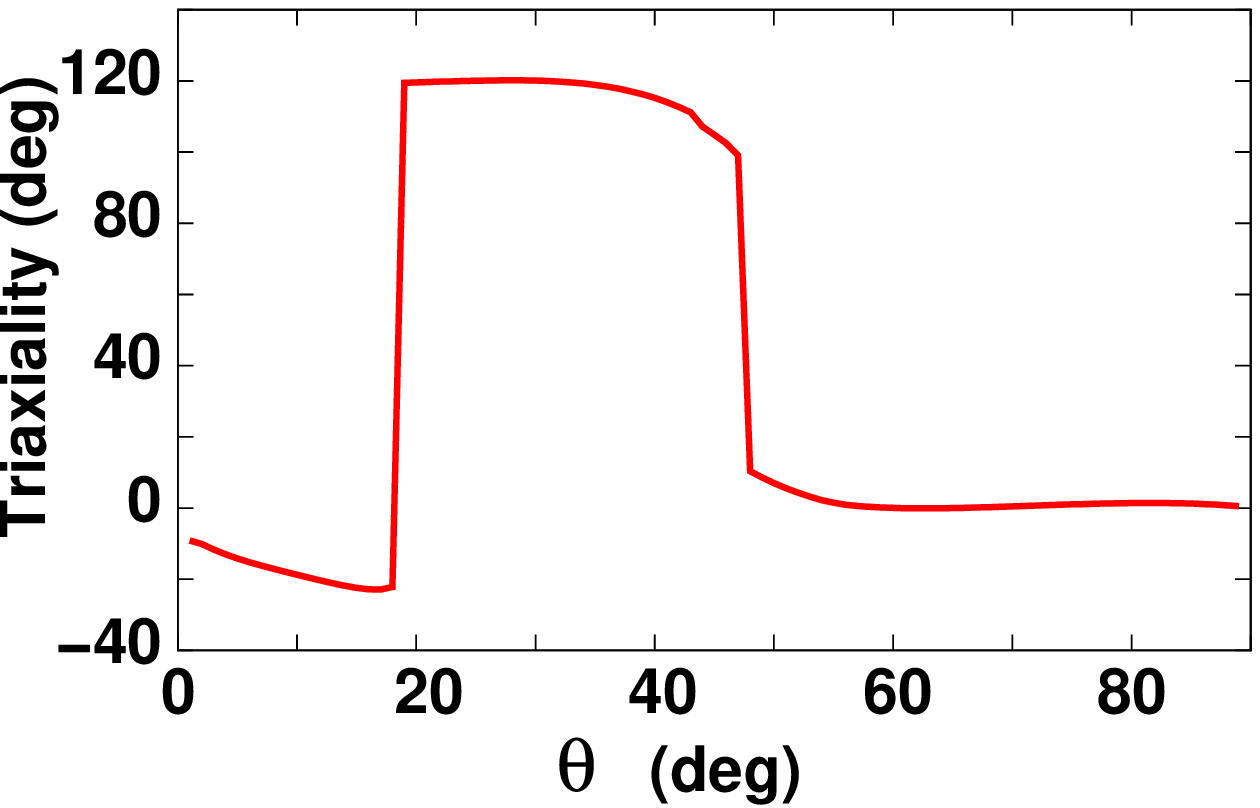}
  \caption{(Top panel) A cross section of the energy surface at $\phi=0^{\circ}$.
  (Bottom panel) Triaxial deformation at $\phi=0^{\circ}$.}
\label{fig-ph0}
\end{figure}

\begin{table}[th]
  \begin{tabular}{ccccccccccc}
    \hline
    $\theta$ \quad \quad &0 & 10 & 20 & 30 & 40 & 50 & 60 & 70 & 80 & 90 \\
    \hline
    $\beta$  \quad \quad &0.31& 0.25 & 0.21 & 0.18& 0.14 & 0.11& 0.17& 0.21& 0.23& 0.24\\
    \hline
  \end{tabular}
  \caption{Change of $\beta$ as a function of the tilt angle $\theta$, obtained
  in the 3D-cranking calculation for $J=8\hbar$ amd $\phi=0^{\circ}$.}
\end{table}

In the lower panel of Fig.\ref{fig-ph0}, the triaxial deformation
is seen to have $\gamma\simeq 120^{\circ}$ at $20^{\circ}\alt \theta \alt 
45^{\circ}$. Because the ``level crossing'' happens at $\theta\simeq 20^{\circ}$,
we cannot exactly see how the graph continues toward $\theta\rightarrow 0$. 
However, from the trend of the graph, 
it is possible to guess that the graph forms a symmetric
shape with respect to $\theta=45^{\circ}$ in the upper panel of Fig.\ref{fig-ph0}
, and that $\gamma\rightarrow 120^{\circ}$ for $\theta\rightarrow 0^{\circ}$
in the lower panel of Fig.\ref{fig-ph0}. This reflection symmetry around
$\theta=45^{\circ}$ indicates that the
state in $0^{\circ}\le \theta \le 45^{\circ}$ and the state 
in $45^{\circ}\le \theta \le 90^{\circ}$ have the same intrinsic structure.

\section{Conclusion}
The ground-state rotational band was studied with the self-consistent
1D-cranking calculation. It was confirmed in this study that 
an effect of triaxiality in the nature of the rotational band 
is important at high spin, as previously anticipated by Valor, et al.
In addition, two high-spin states at $J=8^+$ observed in experiment 
were analysed by means of the self-consistent 3D-cranking calculation.
It concludes (in a  qualitative manner) 
that the yrast $8^+$ state is an axially-symmetric
high-$K$ state created by the deformation-aligned protons and neutrons
in the d$_{5/2}$ orbitals while the second $8^+$ is a rotational member
of the ground-state rotational band with substantial triaxial deformation.

For the first time, in the framework of the self-consistent and
microscopic method, the two $I^{\pi}=8^+$ states in $^{24}$Mg,
which correspond to low- and high-$K$ states respectively,
are explained on the same footing, that is, 
through the self-consistent 3D-cranking model.

\section{acknowledgments}
  The author thanks Professors P. Ring and P.-H. Heenen for their suggestions
  to calculate this $^{24}$Mg nucleus using the self-consistent 3D-cranking 
  method. Useful discussions with Dr W. Catford, 
  Professor P. W. Walker and Dr P. Regan
  are also acknowledged. This work is supported by EPSRC
  with an advanced research fellowship GR/R75557/01 
  as well as a first grant EP/C520521/1.

\end{document}